\documentclass[a4paper,11pt]{article}
\usepackage{pos}
\usepackage{graphicx}
\usepackage{subcaption}
\usepackage{setspace}
\newcommand{\FIXME}[1]{[\textcolor{red}{FIXME: #1}]}

\renewcommand{\hookAfterAbstract}{%
    \par\bigskip
    \textsc{LA-UR-22-33079}
}
\title{Electroweak box diagrams on the lattice for pion and neutron decay}

\author*[a]{Jun-Sik Yoo}
\author[a]{Tanmoy Bhattacharya}
\author[a]{Rajan Gupta}
\author[a,b]{Santanu Mondal}
\author[c]{Boram Yoon}

\affiliation[a]{Los Alamos National Laboratory, Theoretical Division T-2, Los Alamos, New Mexico 87545, USA}

\affiliation[b]{Department of Physics and Astronomy, Michigan State University, MI, 48824, USA}
\affiliation[c]{Los Alamos National Laboratory, Computer Computational and Statistical Sciences Division CCS-7, Los Alamos, New Mexico 87545, USA}

\emailAdd{junsik@lanl.gov}

\abstract{CKM matrix is unitary by construction in the standard model(SM). The recent analyses on the first row of CKM matrix show $ \approx 3\sigma$ tension with unitarity.  Nonperturbative
calculations of the radiative corrections can reduce the theory uncertainty in CKM matrix elements.
Here we compute the electroweak box contribution to the pion and kaon $\beta$ decays using seven $N_f=2+1+1$ HISQ-Clover lattice with various pion mass and lattice spacing.
The continuum and chiral limit is taken using the leading dependence on $M_\pi$ and $a$, where $M_\pi$ extrapolation is taken to the physical pion mass and $SU(3)$ symmetric mass  for pion and kaon box contribution, respectively.
Our results are $ \square_{\gamma W}^{VA} |_{\pi}  = 2.820 (28) \times 10^{-3} $ and $ \square_{\gamma W}^{VA} |_{K} = 2.384 (17) \times 10^{-3} $. 
}

\FullConference{%
  The 39th International Symposium on Lattice Field Theory (Lattice2022),\\
  8-13 August, 2022 \\
  Bonn, Germany 
}

\begin{document}
\maketitle

\section{Introduction}
In the precision frontier, physics beyond the standard model (BSM) is probed by confronting accurate predictions of the standard model (SM) with precision experiments. Today, there are several tests showing roughly 2-3$\sigma$ deviations, one being the unitarity of the first row of the CKM quark mixing matrix: $\Delta_{\textrm{CKM}} \equiv\allowbreak |V_{ud}|^2+\allowbreak|V_{us}|^2+\allowbreak|V_{ub}|^2 - 1$ should be zero. Current analyses show  a {$\approx 3\sigma$} tension with the SM~\cite{Workman:2022ynf,Seng:2018yzq,Seng:2018qru,Czarnecki:2019mwq} using the most precise value of $|V_{ud}|^2=0.94815(60)$ coming from $0^+\!\! \to  0^+ $   nuclear $\beta$ decays \cite{Workman:2022ynf}, while $|V_{us}|^2=0.04976(25)$ is obtained from kaon semileptonic decays ($K \to \pi \ell \nu_\ell$) with the $N_f = 2+1+1$ lattice result for $f^K_+(0)$~\cite{Aoki:2021kgd}, and   $|V_{ub}|^2\approx(2\pm 0.4) \times 10^{-5}$ has no impact on the unitarity test.

\begin{figure}
    \centering
    \includegraphics[width=.6\textwidth]{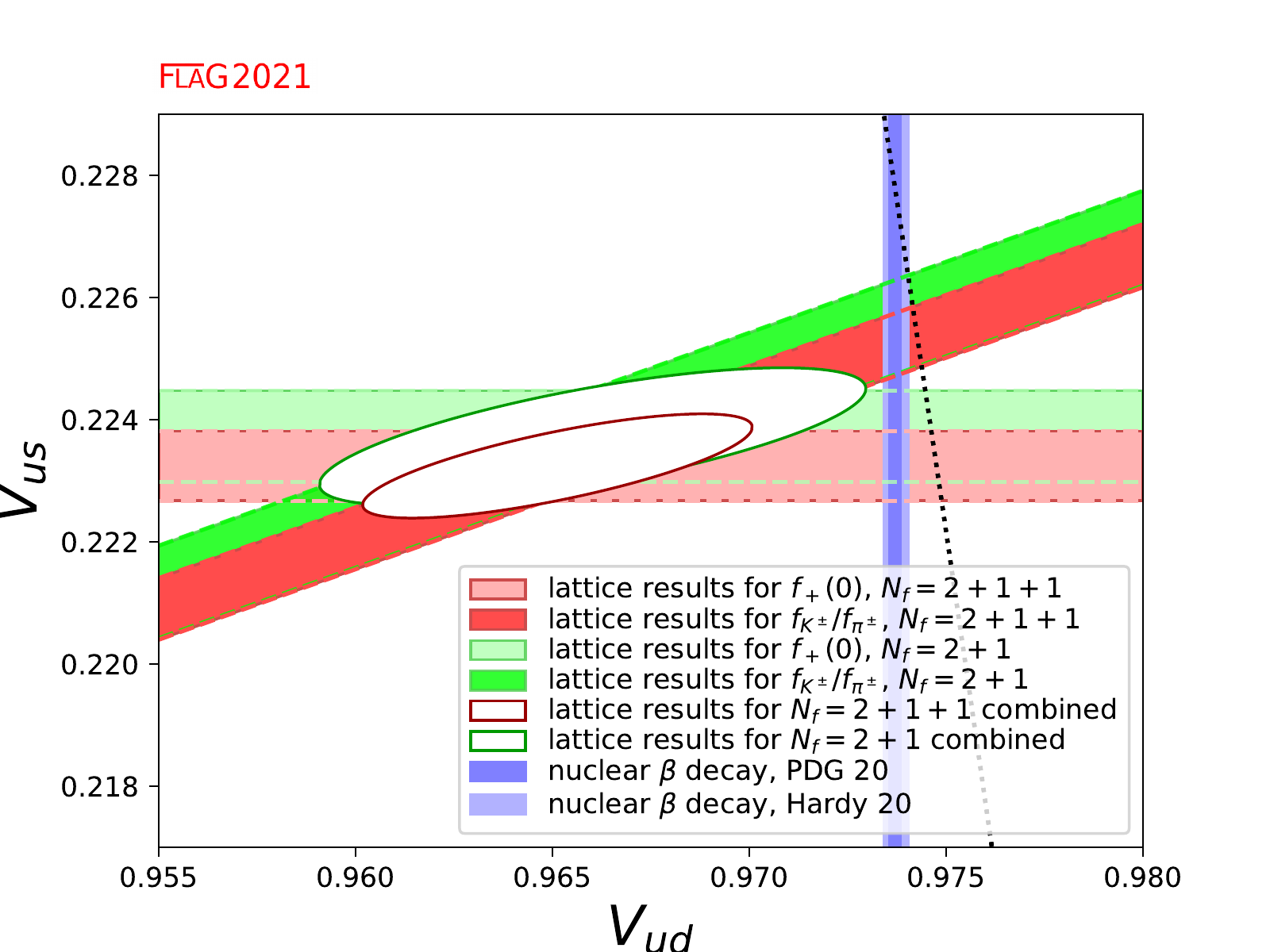}
    \caption{Theory/Experimental bounds on $V_{ud}$ and $V_{us}$  in the first row of the CKM matrix\cite{FlavourLatticeAveragingGroupFLAG:2021npn} }
    \label{fig:FLAG20}
\end{figure}

\begin{figure}
    \centering
    \includegraphics[width=.8\textwidth]{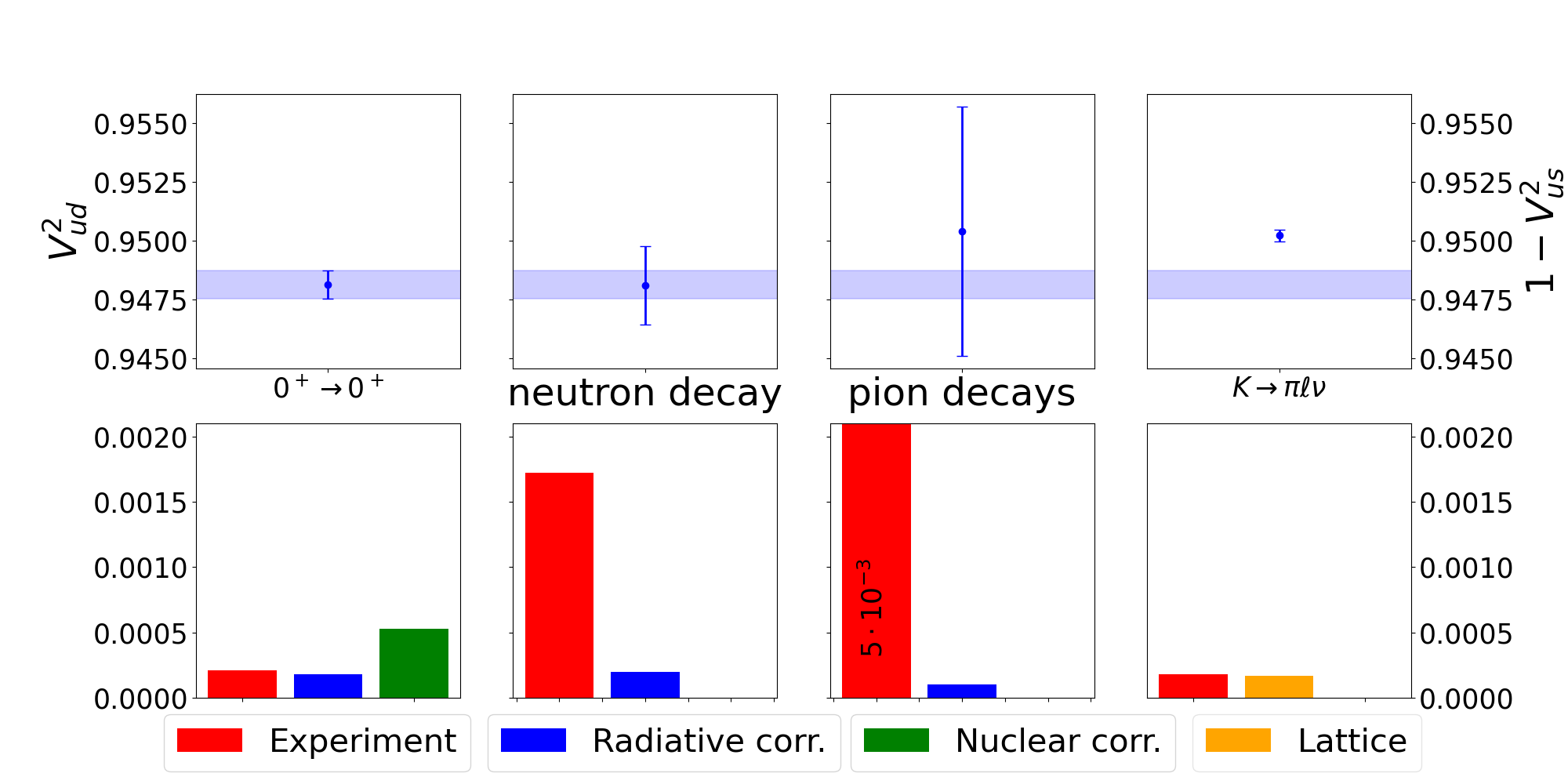}
    \caption{Error budget on the first row components of CKM matrix \cite{Workman:2022ynf,UCNt:2021pcg,Hardy_2012}. }
    \label{fig:Errorbudget}
\end{figure}

A current analysis of the unitarity bound is shown in Fig.~\ref{fig:FLAG20}, with the error budget in Fig.~\ref{fig:Errorbudget}. The extraction of $V_{ud}$ from superallowed $0^+ \rightarrow 0^+$ nuclear decays is the best, however, there is significant uncertainty in the theoretical analysis of nuclear effects. The goal of our lattice calculations is to provide a controlled estimate of the non-perturbative region of the electroweak $\gamma W$-box diagram (Fig.~\ref{fig:diagrams} left) needed to reduce the uncertainty in the radiative corrections (RC) to neutron decay \cite{Sirlin:1977sv}, which together with improvements in experiments will make the extraction of $V_{ud}$ from it competitive. \looseness-1

$|V_{ud}|^2$ from neutron decay is given by the
master formula~\cite{Czarnecki:2019mwq,Czarnecki:2018okw}
\vspace{-6pt}\begin{eqnarray}\label{master}
\left|V_{ud}\right|^2 = \left(  \frac{G_\mu^2 m_e^5 }{2\pi^3} f \right)^{-1} \frac{1}{ \tau_n (1 + 3 g_A^2) (1 + \textrm{RC}) }
= \frac{5099.3(3) \textrm{s}}{\tau_n (1 + 3 g_A^2) (1 + \textrm{RC})}
\end{eqnarray}
where $\tau_n$ is the free neutron lifetime, $g_A$ is the axial coupling, which can be
obtained from the neutron $\beta$ decay asymmetry parameter $A$,
$G_\mu$ is the Fermi constant extracted from muon decays, and
$f=1.6887(1)$ is a phase space factor. With future measurements of the neutron lifetime $\tau_n$ reaching an uncertainty of $\Delta\tau_n \sim 0.1$~s, and of ratio $\lambda = g_A/g_V$ of the neutron axial and vector coupling reaching  $\Delta\lambda/|\lambda|\sim 0.01\%$, the extraction of $V_{ud}$ with accuracy comparable to $0^+\!\! \to  0^+ $ superallowed $\beta$ decay can be achieved provided the uncertainty in the RC to neutron decay can be reduced. 

So far we have results for RC to pion and kaon decays and are working on methods to get a signal in the neutron correlation functions. For pion and kaon semileptonic decays, the analogues of Eq.~\eqref{master} to extract $|V_{ud}|$ and $|V_{us}|$ are 
\begin{equation}
    |V_{ud}f_+^\pi(0)|_{\pi\ell} = \sqrt{\frac{64\pi^3  \Gamma_\pi}{G_\mu^2 M_\pi^5 I_{\pi} \left( 1 + \delta  \right) }}
\end{equation}\begin{equation}
    |V_{us}f_+^K(0)|_{K\ell} = \sqrt{\frac{192\pi^3 \textrm{BR}(K\ell ) \Gamma_K}{G_\mu^2 M_K^5 C_K^2 S_{EW} I_{K\ell} \left( 1 + \delta^{K\ell}_{EM} + \delta^{K\ell}_{SU(2)} \right) }} \,,
\end{equation}
where $\Gamma_{\pi/K}$ are $\pi$ and K decay rates, $I_{\pi,K}$ are known kinematic factor, $f_+^{\pi/K}$ are semileptonic form factors, $C_K$ is a normalization for kaon decay, $S_{EW}$ is the short distance radiative correction, $\delta$ is RC to $\pi$ decay, $\delta^{K\ell}_{EM}$ is the long distance correction, and the $\delta^{K\ell}_{SU(2)} $ is the isospin breaking correction.

\section{Lattice Setup}
The parameters of the seven $N_f = 2+1+1$ HISQ sea quark ensembles, generated by the MILC collaboration~\cite{Bazavov:2012xda}, are given in Table~\ref{tab:my_label}. The correlation functions are constructed using the clover action as described in Ref.~\cite{Gupta:2018qil}. The strong coupling $\alpha_s$ at each lattice ensemble was computed up to fourth order \cite{Deur:2016tte}. The calculation of the hadronic tensor with the insertion of vector (V) and axial (A) currents gives rise to 4 types of Wick contractions. These quark-line diagrams are shown in Fig.~\ref{subfig:diagrams}, with (B) absent for the nucleon.
The relevant connected diagrams are A and C. The disconnected diagram (B) does not contribute due to the $\gamma_5-$hermiticity property of the quark propagator, and diagram (D) vanishes in the SU(3) limit and is not evaluated here. 
We generate quark propagators with wall sources at the two ends of a sublattice with separation $\tau$ and label these quark lines by W.
For the internal line S in diagram C, we solve for an additional propagator from the position of the vector current $V_\mu$ on the timeslice in the middle between the source and sink. This point is labeled $\{{\vec x}=0, t=0\}$, and we choose 256 points for diagram A and 64 for diagram C. Data are collected with the position of $A_\mu$ varied within distance $R^2$ from these points. 
On each configuration, we use 8 regions (sublattices) offset by $N_T / 8$ on which we repeat the calculation to further increase the statistics. 

\begin{table}[]
    \centering
    \begin{tabular}{c|c|c|c|c|c|c|c|c}
    \hline \hline 
        Ensemble ID & a[fm] & $\alpha_S$ & $m_\pi^{val}$[MeV] & $L^3 \times T$ & $m_\pi L$ & $\tau / a$ & $R^2$ & $N_{conf}$  \\
        \hline 
        a06m310 & .0582(04) & .2580 &319.3(5) & $48^3 \times 144$ & 4.52 & 46 & 1600 & 168 \\
        \hline
        a09m130 & .0871(06) & .3087 &138.1(1.0) & $64^3 \times 96$ & 3.90 & 40 & 800 & 45 \\
        a09m310 & .0888(08) & .3117&313.0(2.8) & $32^3 \times 96$ & 4.51 & 40 & 400 & 156 \\
        \hline 
        a12m220 & .1184(09) & .3660 &227.6(1.7) & $32^3 \times 64$ & 4.38 & 18 & 400 & 99  \\
        a12m220L & .1189(09) & .3660 &227.9(1.9) & $40^3 \times 64$ & 5.49 & 30 & 400 & 50  \\
        a12m310 & .1207(11) & .3704& 310.2(2.8) & $24^3 \times 64$ & 4.55 & 18 & 400 & 179  \\
        \hline 
        a15m310 & .1510(20) & .433&320.6(4.3) & $16^3 \times 48$ & 3.93 & 24 & 400 & 80  \\
        \hline \hline 
    \end{tabular}
    \caption{The 7 HISQ-Clover Lattice ensembles used in this work. On each configuration, we use 8 sublattice regions and in each make 256 measurements for diagram A and 64 for diagram C.}
    \label{tab:my_label}
\end{table}

\section{Electroweak Box Diagram}
\label{sec:EWbox}

\begin{figure}
\begin{subfigure}{0.39\textwidth}
\includegraphics[width=\linewidth]{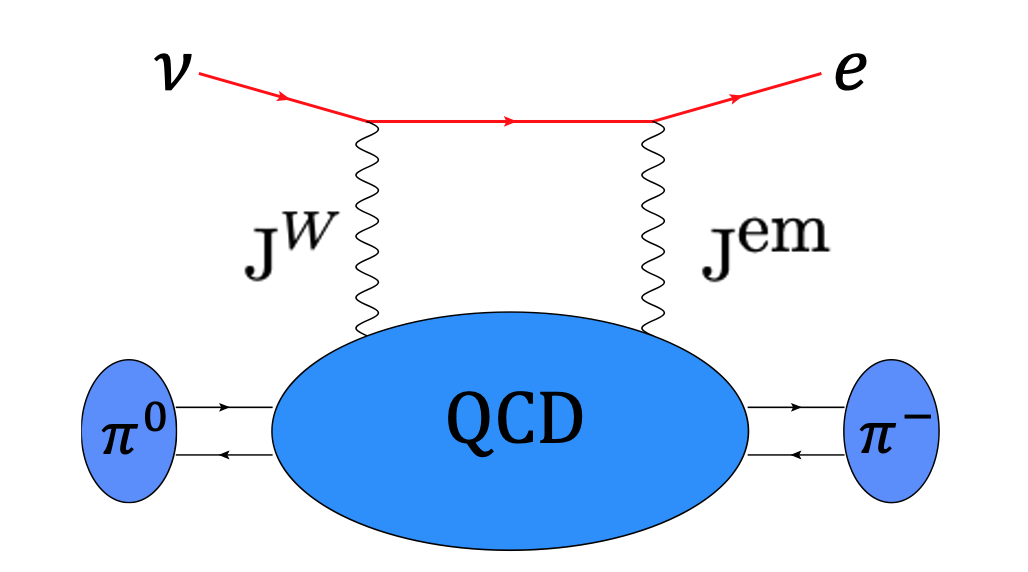}
\caption{Pion box diagram} \label{subfig:feynman_diag}
\end{subfigure}
\hspace*{\fill}
\begin{subfigure}{0.60\textwidth}
    (A)\includegraphics[width=0.33\linewidth]{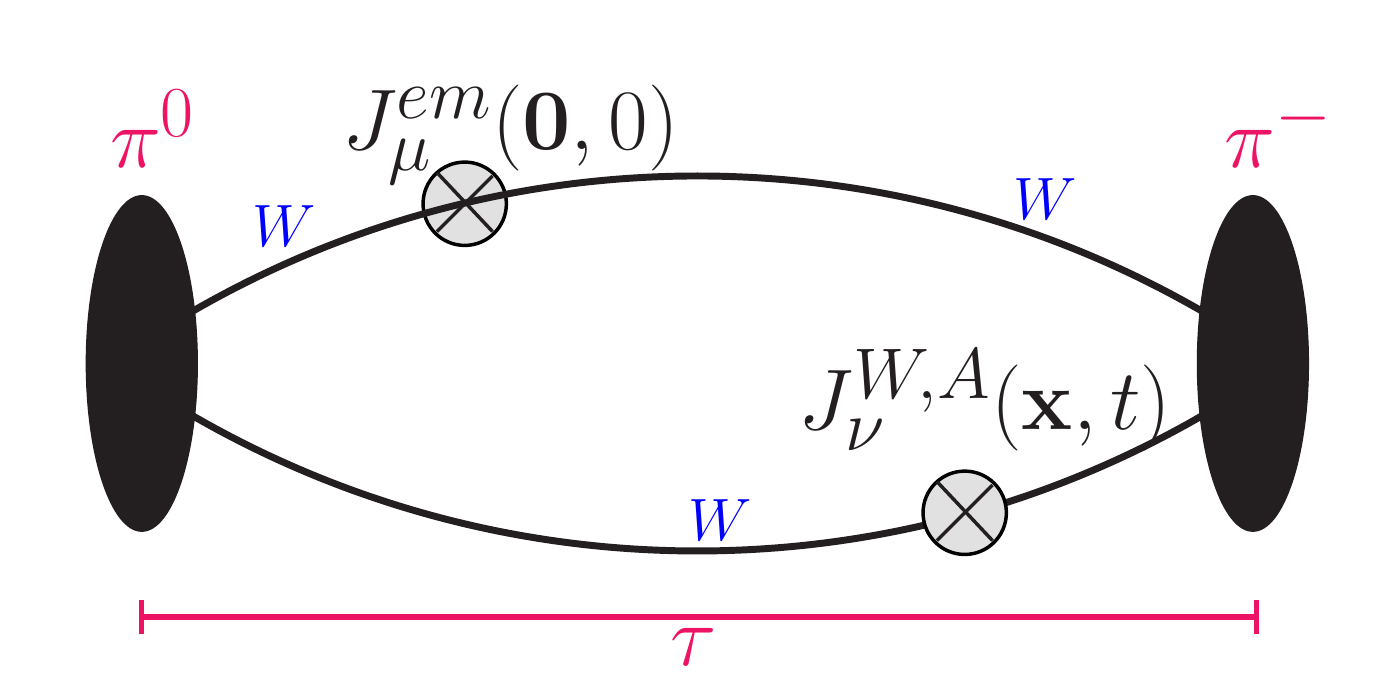} \qquad
    (B)\includegraphics[width=0.33\linewidth]{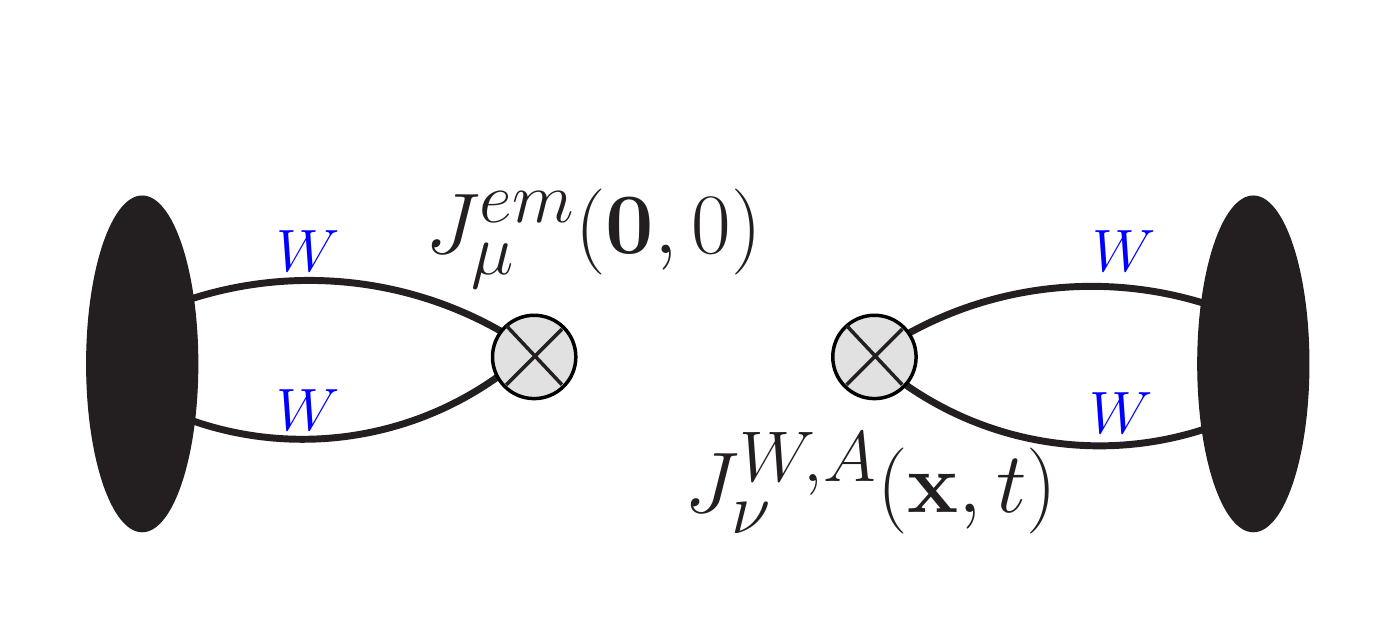} \\
    (C)\includegraphics[width=0.33\linewidth]{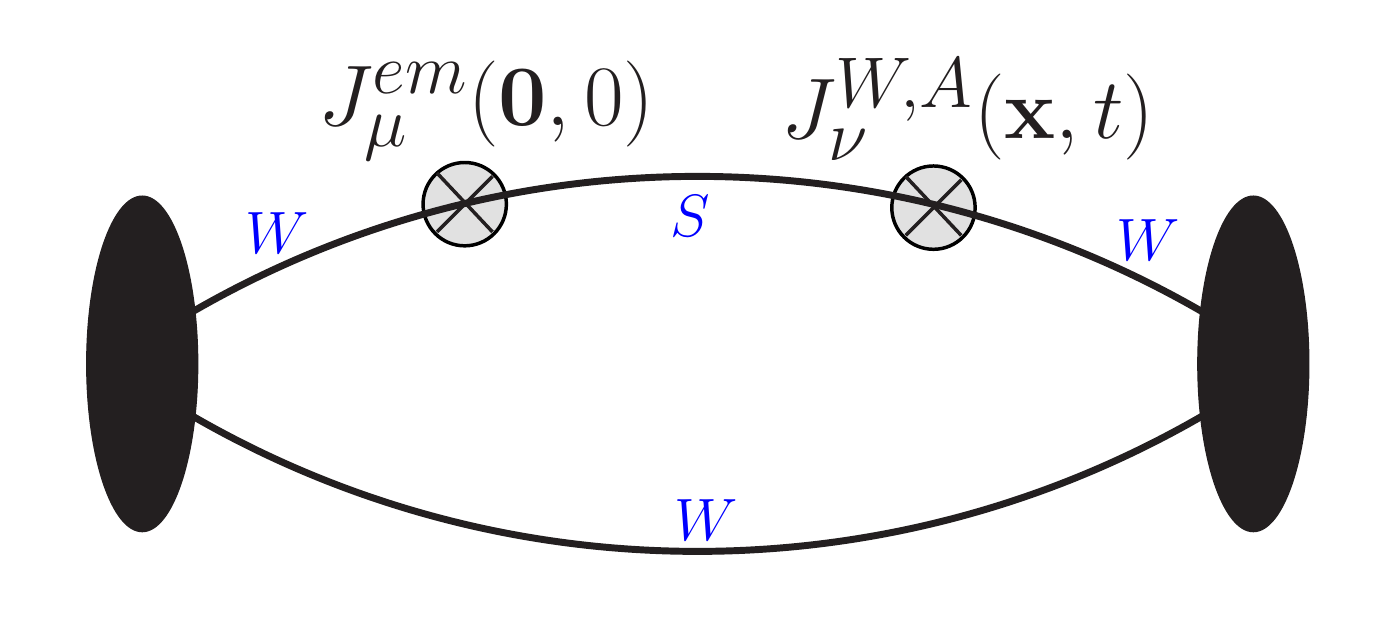} \qquad
    (D)\includegraphics[width=0.33\linewidth]{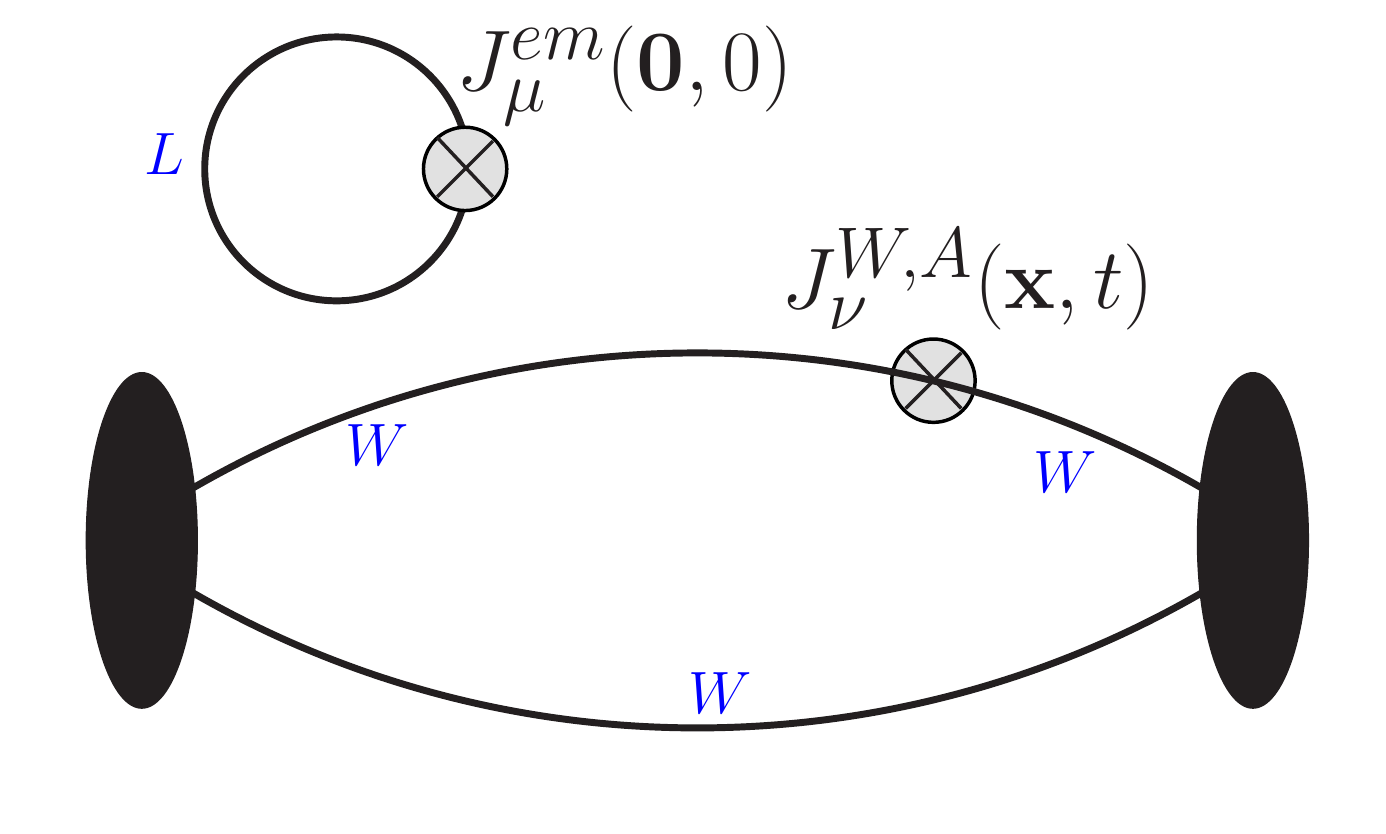}
\caption{quark-line diagrams} \label{subfig:diagrams}
\end{subfigure}
\caption{Axial $\gamma W-$ box diagram for RC to pion decay (left), and the 4 quark-line diagrams that contribute to the pion $\gamma W$-box $\mathcal{H}_{\mu \nu}^{V A}(\vec{x},t) = \langle \pi | \textrm{T} [J_\mu^{em}(x) J_\nu^{W,A}(0)] | \pi \rangle $ (right).}
\label{fig:diagrams}
\end{figure}

\begin{figure}[h]
    \centering
    \includegraphics[width=.48\textwidth]{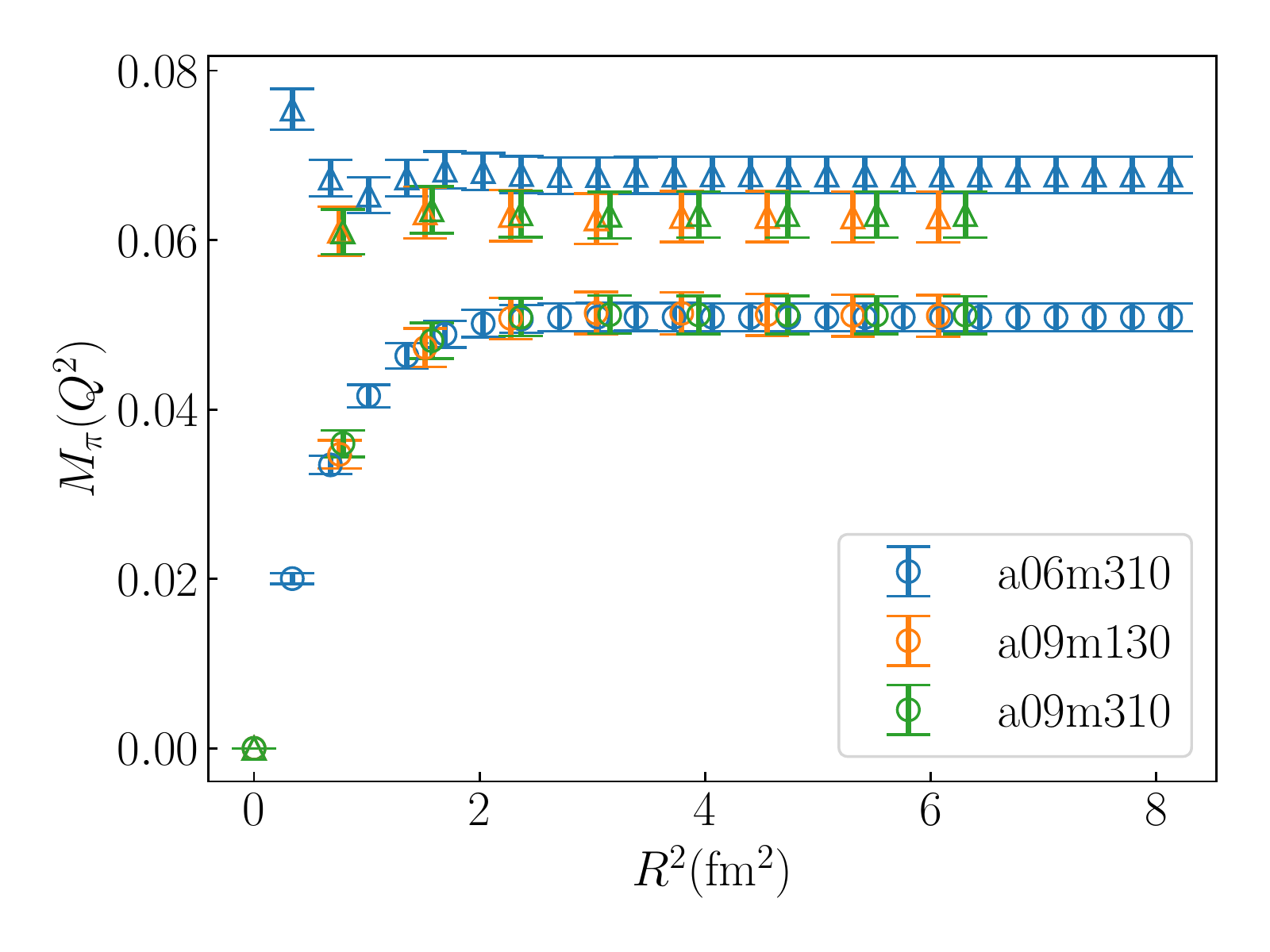}%
    \includegraphics[width=.48\textwidth]{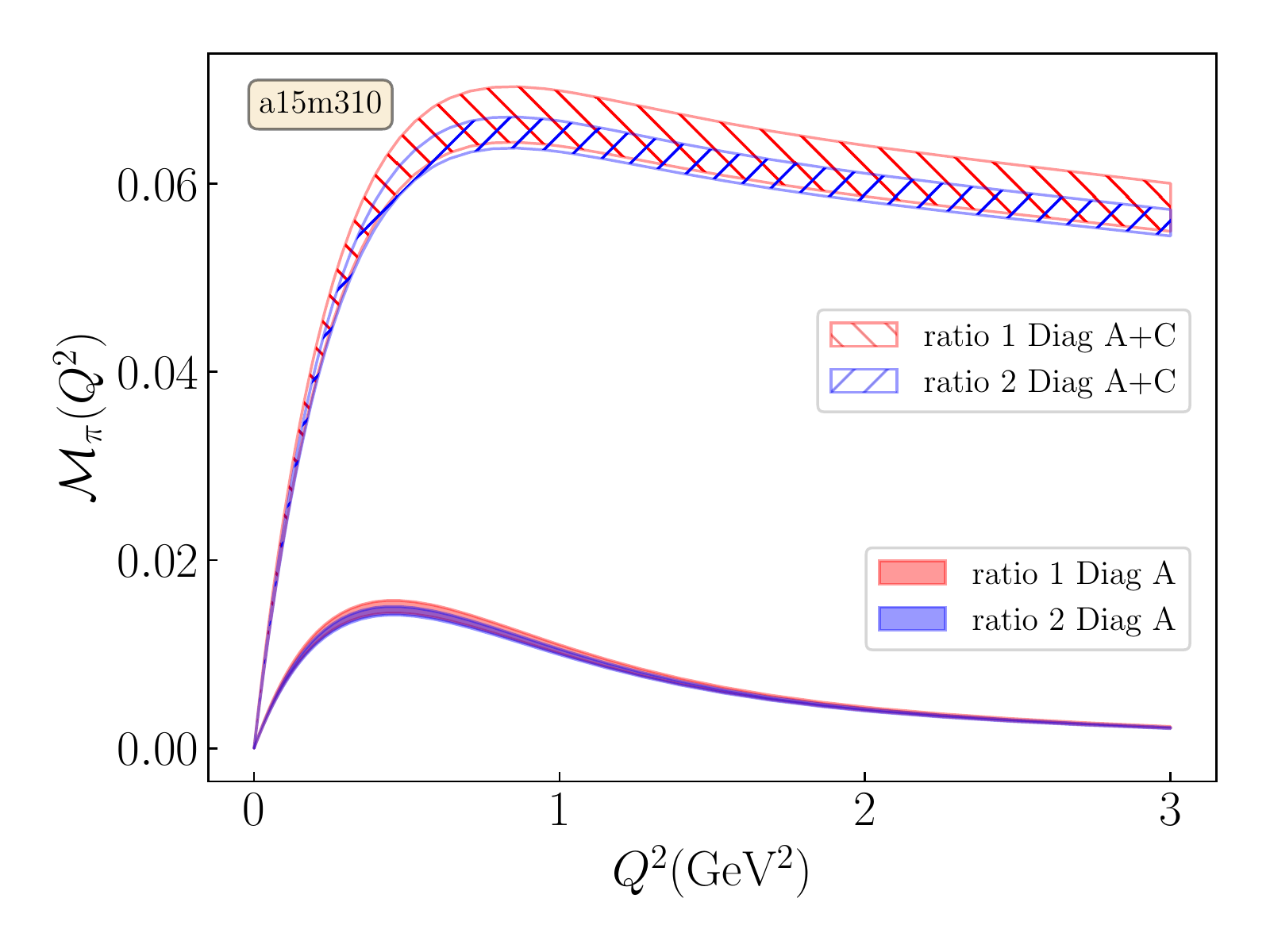}
    \vspace{-0.2cm}
    \caption{(Left) The dependence of ${\cal M}_\pi$ on $R^2$ to check convergence. (See Section~\ref{sec:EWbox} for details.) Circles are used for $Q^2 = 0.317$ GeV$^2$ and triangles for $Q^2 = 3.0$ GeV$^2$ data. (Right) Comparison of the signal in ${\cal M}_\pi (Q^2)$ extracted by (i) ratio 1 combining the ratio defined in Eq.~\eqref{eq:R4to2} and $F_+^{\pi} = \sqrt{2}$ (red), and (ii) ratio 2 using the ratio in Eq.~\eqref{eq:R4to3} (blue). There is roughly a factor of two reduction in errors using Eq.~\eqref{eq:R4to3} as shown by the blue band.\looseness-1}
    \label{fig:Rsq-and-ratio}
\end{figure}

The electroweak box diagram  (called the axial $\gamma W$ diagram), shown in Fig.~\ref{fig:diagrams} (left), is given by~\cite{Feng:2020zdc} (the renormalized currents used are $J^{W,A}_\mu = Z_A \bar{u} \gamma_\mu \gamma_5 d $ and $J_\mu^{em} = Z_V ( \frac{2}{3}\bar{u}  \gamma_\mu d - \frac{1}{3}\bar{d}  \gamma_\mu d)$ with $Z_A$ and $Z_V$ taken from Ref.~\cite{Gupta:2018qil}).

\begin{equation}\label{delta}
    \Delta_{\textrm{}} = \int_0^{+\infty} dQ^2
    \int^{ Q}_{-  Q}  
     d Q_0 \frac{1}{Q^4} \frac{1}{Q^2 + m_W^2} L^{\mu\nu}(Q,Q_0) T^{VA}_{\mu\nu}(Q,Q_0)\ 
\end{equation}
with the relevant hadronic tensor $T^{VA}_{\mu\nu}$ given by 
\begin{equation}
T^{VA}_{\mu\nu}=\frac{1}{2}\int d^4x\,e^{iQ \cdot x}\langle
H_f (p)|T\left[J^{em}_\mu(0,0)J^{W,A}_\nu({\vec x},t)\right]|H_i(p)\rangle \,,
\label{eq:Tmunu}
\end{equation}
with $H$ standing for $\pi,\ K, \ N$ states.   
The spin-independent part of $T^{VA}_{\mu\nu}$ has only one term $T_3$ from
$T^{VA}_{\mu\nu}= i \epsilon_{\mu \nu \alpha \beta} q^\alpha p^\beta T_3 +
\dots$
 Knowing $T_3$ as a function of $Q^2$, the $\gamma W$-box correction is given by
\begin{equation}
\Box^{VA}_{\gamma W} = \frac{3 \alpha_e}{2 \pi} \int \frac{dQ^2}{Q^2}
\frac{M_W^2}{M_W^2+Q^2} {\mathcal{ M}_H(Q^2)} \quad {\rm with} 
\label{eq:boxMH}
\end{equation}
\begin{equation}
    {\mathcal{M}_{H}\left(Q^{2}\right)}=-\frac{1}{6} \frac{1}{F_{+}^{H}} \frac{\sqrt{Q^{2}}}{M_{H}} \int d^{4} x \omega(\vec{x},t) \epsilon_{\mu \nu \alpha 0} x_{\alpha} \mathcal{H}_{\mu \nu}^{V A}(\vec{x},t) \,
\end{equation}
where $M_W$ is the $W$ meson mass, $M_H$ is the hadron mass, $\omega(\vec{x},t)$ is a 
weight function defined in~\cite{Feng:2020zdc}, and $\mathcal{H}_{\mu \nu}^{V A}(\vec{x},t)= \langle \pi | \textrm{T} [J_\mu^{em}(x) J_\nu^{W,A}(0)] | \pi \rangle $ is given by the sum of the four quark-line diagrams (for mesons) in Fig.~\ref{subfig:diagrams} (right).  $\mathcal{H}_{\mu \nu}^{V A}(\vec{x},t)$ is a function of the separation $\{\vec x, t\}$, and on the lattice, the integral becomes a sum, however, ${\mathcal{M}_{H}\left(Q^{2}\right)}$ is available for all values of $Q^2$. 
We expect the signal in $\mathcal{H}_{\mu \nu}^{V A}(\vec{x},t)$ to fall off with $\{\vec x, t\}$, therefore summing over a finite region of radius $R$ should suffice. In Fig.~\ref{fig:Rsq-and-ratio} (Left), we show that the integral saturates for $ R^2 \gtrsim 2.0 \textrm{fm}^2$. To save computation time, but stay on the conservative side, we choose the integration volume to be larger than $ R^2 \sim 3.3 \textrm{fm}^2$.

\vspace{-.5cm}
\section{Error reduction in the extraction of $\mathcal{H}_{\mu \nu}^{V A}$}
\label{sec:err_red}
The spectral decomposition of the two-point correlator of the pion is:
\begin{align}
    C_{2pt}(\tau) &= C^{fwd}_{2pt}(\tau) + C^{bkw}_{2pt}(\tau) = \sum_{\mathbf{x}} e^{-i\mathbf{p}\cdot \mathbf{x}}\langle J_\pi(\tau,\mathbf{x}) J_\pi^\dag(0) \rangle \\
    &= \sum_{i} \Big| \langle 0 | J_\pi | \pi_i (p) \rangle \Big|^2 \frac{e^{-E_i(\mathbf{p})\tau } + e^{-E_i(\mathbf{p})( T-\tau )}}{2E_i(\mathbf{p})} 
\end{align} 
where $i$ indexes the excited states. Statistics for $C_{2pt}(\tau)$ is  increased by averaging over forward and backward propagation. 

The spectral decomposition of the hadronic tensor, limited to zero momentum source and sink by using wall sources for quark propagators,  and normalized by the 2-point function, is 
\begin{align}
    R^H_{\mu \nu} (t, \tau, \vec{x}) 
    &= \frac{ C_{4pt}(\tau, t, \vec{x})}{C^{fwd}_{2pt}(\tau)}
    = \frac{ 2M_\pi \langle J_{\pi^0}(\tau/2) J^{em}_\mu(0,0)J^{W,A}_\nu({\vec x},t) J_{\pi^{-}}(-\tau/2)\rangle}{ \Big| \langle 0 | J_\pi | \pi \rangle \Big|^2 e^{-M_\pi\tau } },
    \label{eq:R4to2}
\end{align}
where for $C^{fwd}_{2pt}(\tau)$ one can use the fit or the data. Note that for pseudoscalar mesons we can truncate to just the ground state contribution since the ($V,A$) insertions can both be made in the plateau region, i.e., far enough away from both source and sink timeslices to kill excited states. 
\begin{align}
    \lim_{\tau \rightarrow \infty} R^H_{\mu \nu} (t, \tau, \vec{x}) 
    &= \langle \pi^0(p)|T\left[J^{em}_\mu(0,0)J^{W,A}_\nu({\vec x},t)\right]|\pi^{-}(p)\rangle / 2M_\pi 
    = \mathcal{H}_{\mu \nu} / 2M_\pi
\end{align}
The form factor $F^H_+ $ (matrix element) is obtained from the 3-point function,
\begin{equation}
    F^H_+ = \frac{\langle H(p') |J^{W,V}_\mu | H(p) \rangle }{(p+p')_{\mu=4}} =\frac{\sqrt{2}C_{3pt}(\tau)}{ C^{fwd}_{2pt}(\tau)},
     \label{eq:R3to2}
\end{equation}
for $H=\pi$. for $H=K$, the factor $\sqrt 2$ is absent.  Thus, we can calculate the desired ratios 
\begin{align}
    \frac{\mathcal{H}_{\mu \nu}^{V A}(t, \vec{x})}{F_{+}^{\pi}} &= 2M_\pi \frac{C_{4pt}(\tau, t, \vec{x})}{\sqrt{2}C_{3pt}(\tau)} \quad 
    {\rm and} \quad
    \frac{\mathcal{H}_{\mu \nu}^{V A}(t, \vec{x})}{F_{+}^{K}}
    = 2M_K \frac{C_{4pt}(\tau, t, \vec{x})}{C_{3pt}(\tau)}
     \label{eq:R4to3}
\end{align}
in two ways. Using the left hand side with  $(F_+^\pi(0),F_+^K(0)) = (\sqrt{2}, 1)$ (including normalization factors) or as the ratio of correlation functions. As shown in Fig.~\ref{fig:Rsq-and-ratio} (Right), there is larger cancellation of correlations between the 3- and 4-pt functions, so we exploit the second method. 

\section{Comparing lattice results for ${\cal M}_H(Q^2)$ with perturbation theory}

\begin{figure}
    \centering
    \includegraphics[width=.5\textwidth]{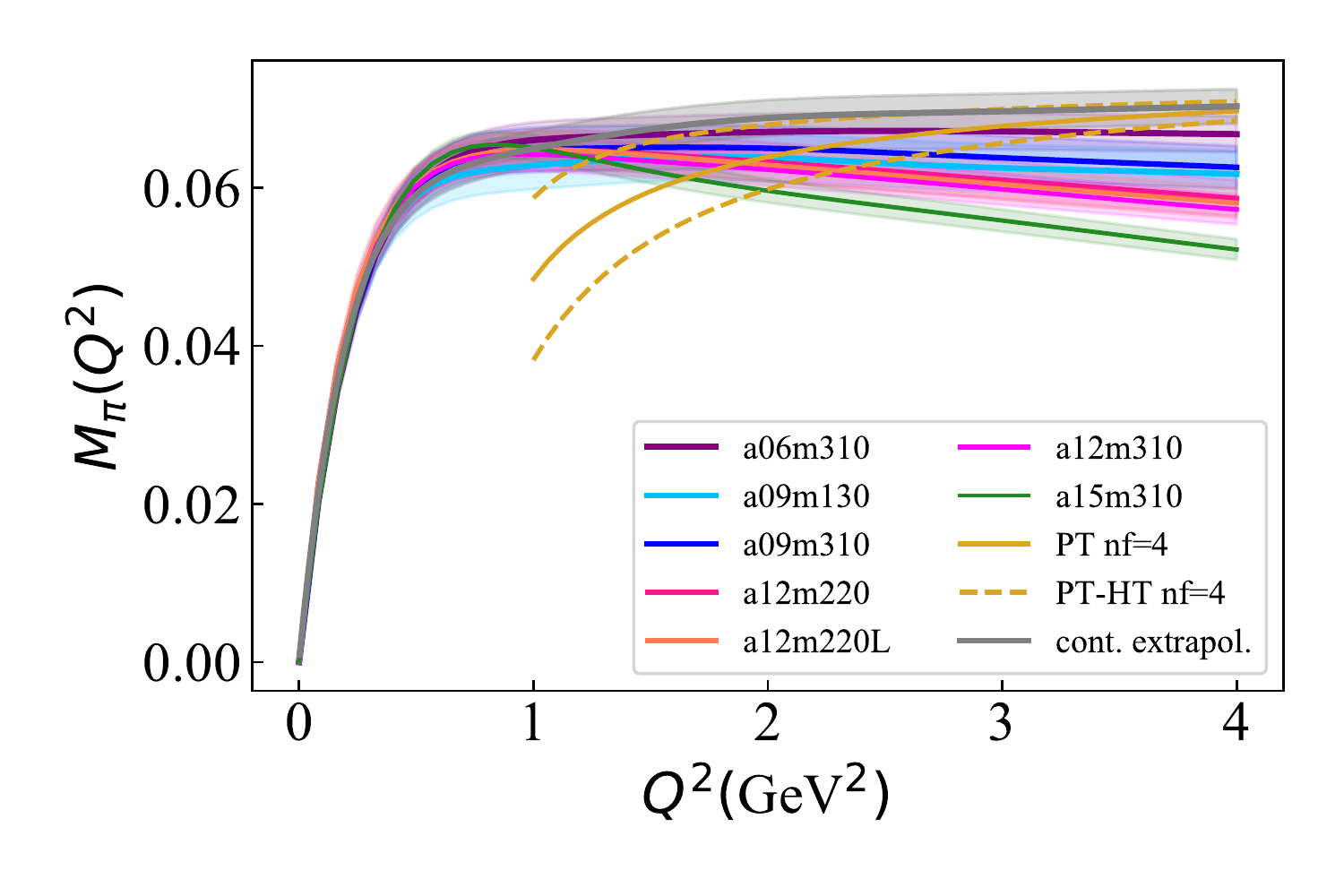}%
    \includegraphics[width=.5\textwidth]{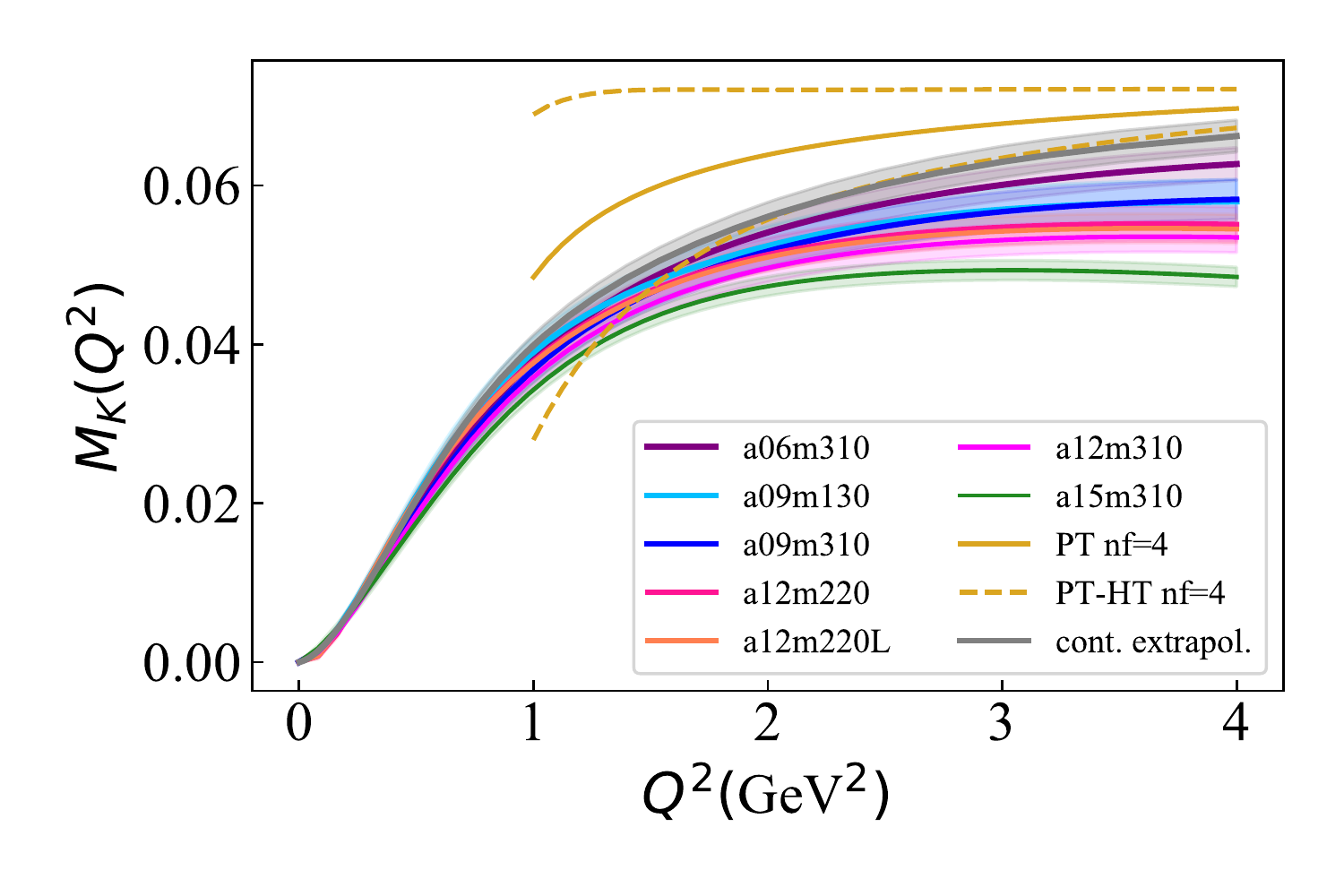}
    \vspace{-.5cm}\includegraphics[width=.5\textwidth]{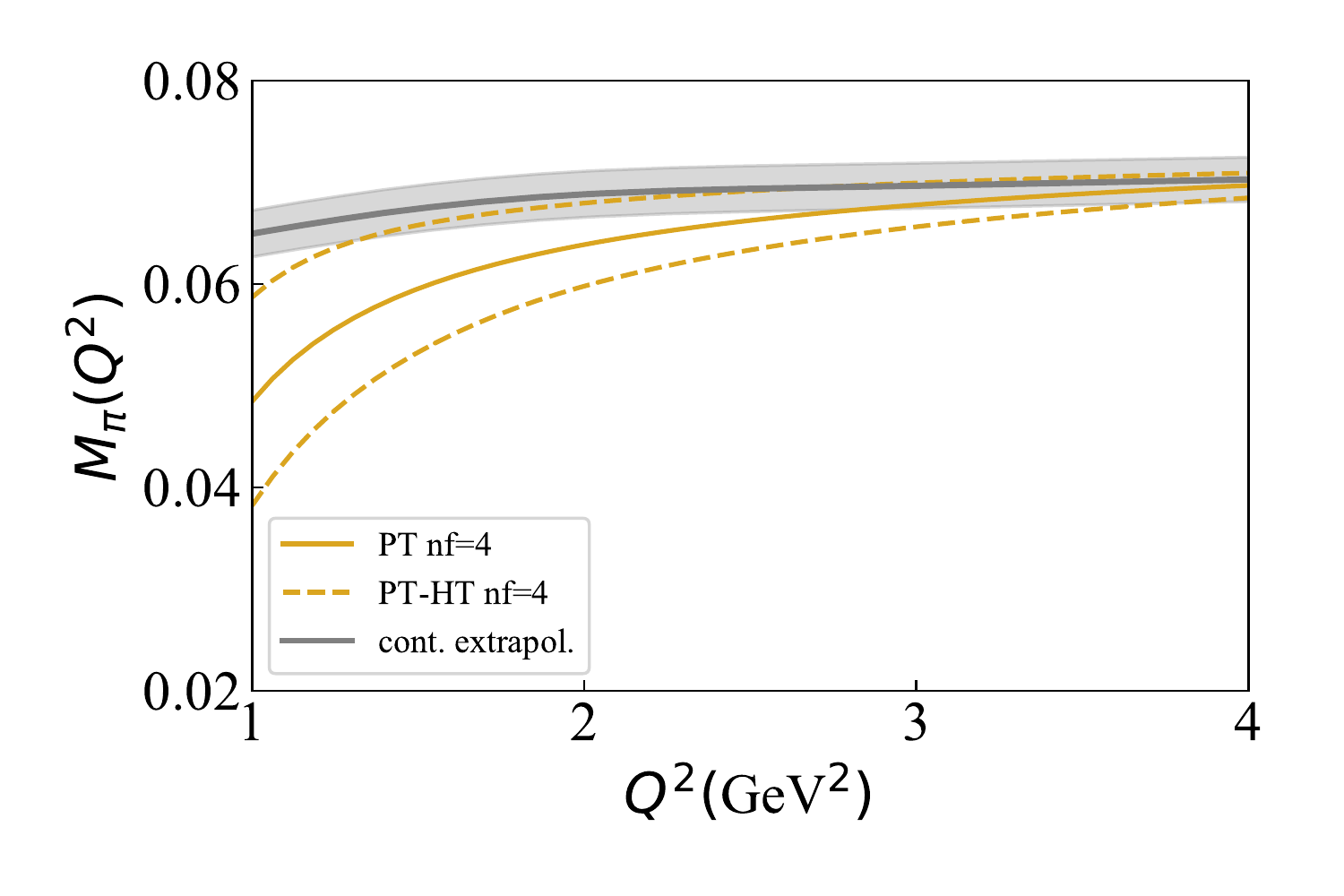}%
    \includegraphics[width=.5\textwidth]{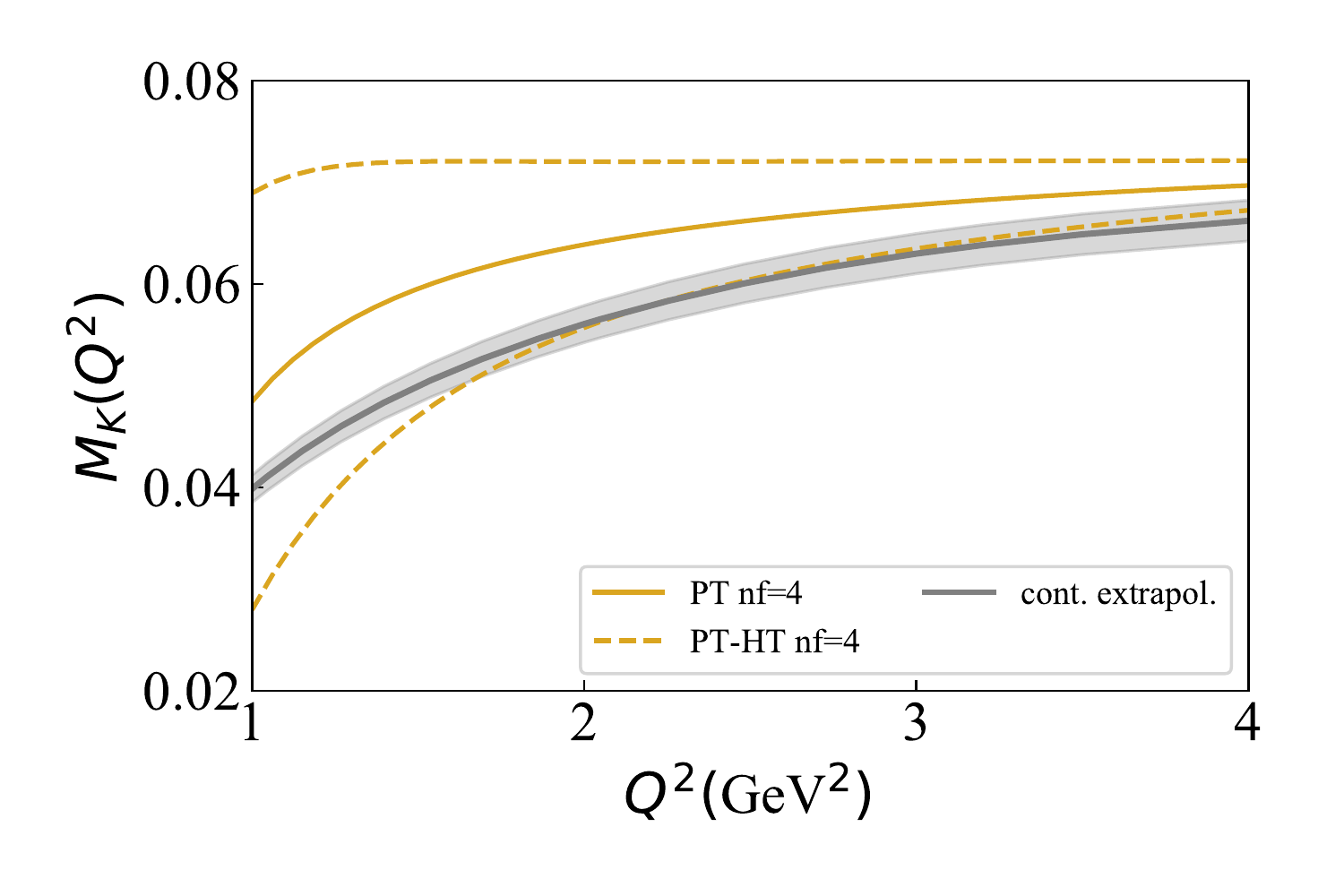}
    \caption{$M_H(Q^2)$ for the (a) pion and (b) the Kaon for the 7 ensembles (top). The bottom panels zoom in on the comparison between the grey band obtained by making a continuum extrapolation at each of the 60 $Q^2$ values and the gold line shows the perturbative result with uncertainty band reflecting higher-twist corrections.\looseness-1}
    \label{fig:M_H}
\end{figure}

\begin{figure}
    \centering
    \includegraphics[width=.5\textwidth]{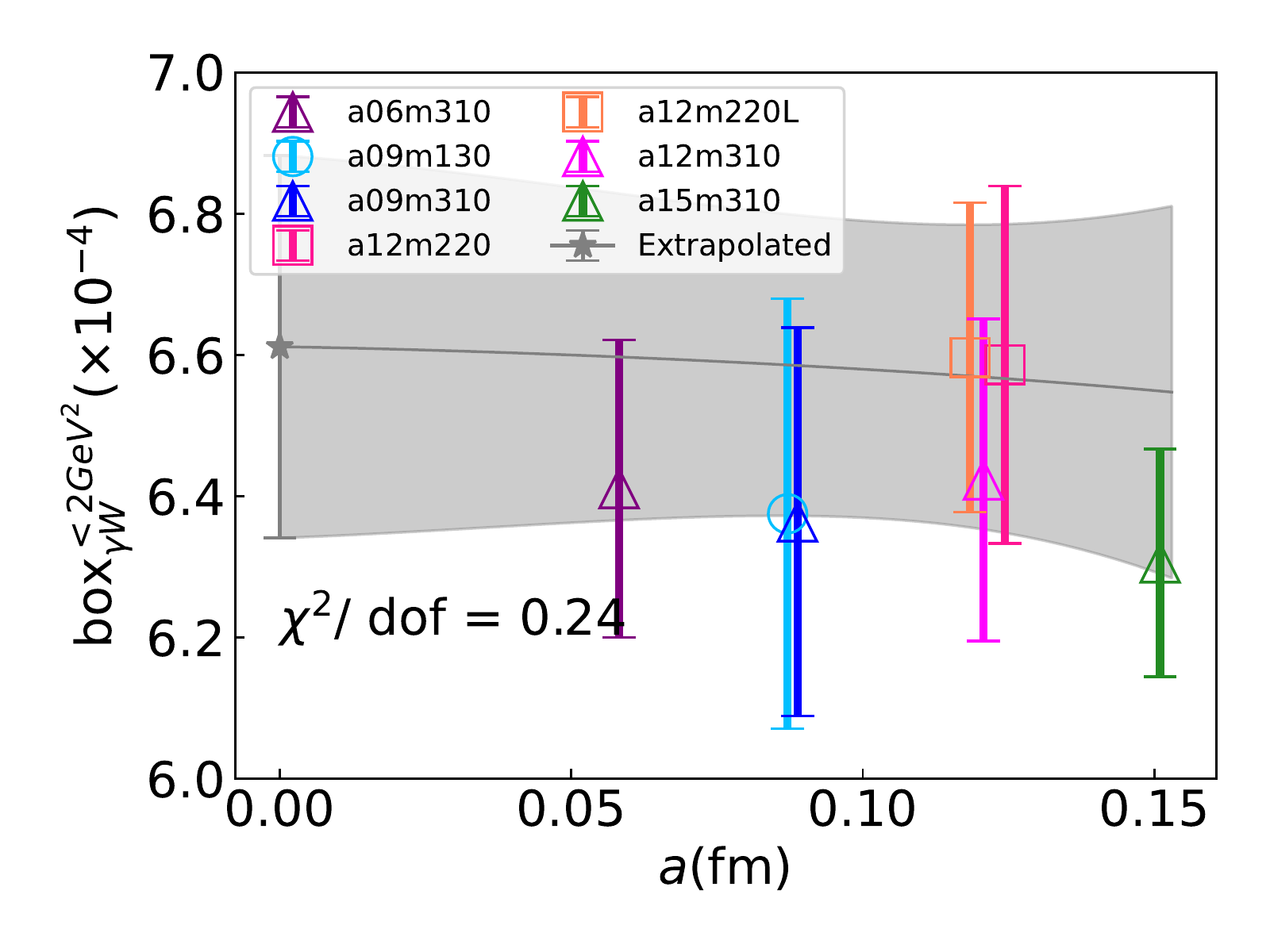}%
    \includegraphics[width=.5\textwidth]{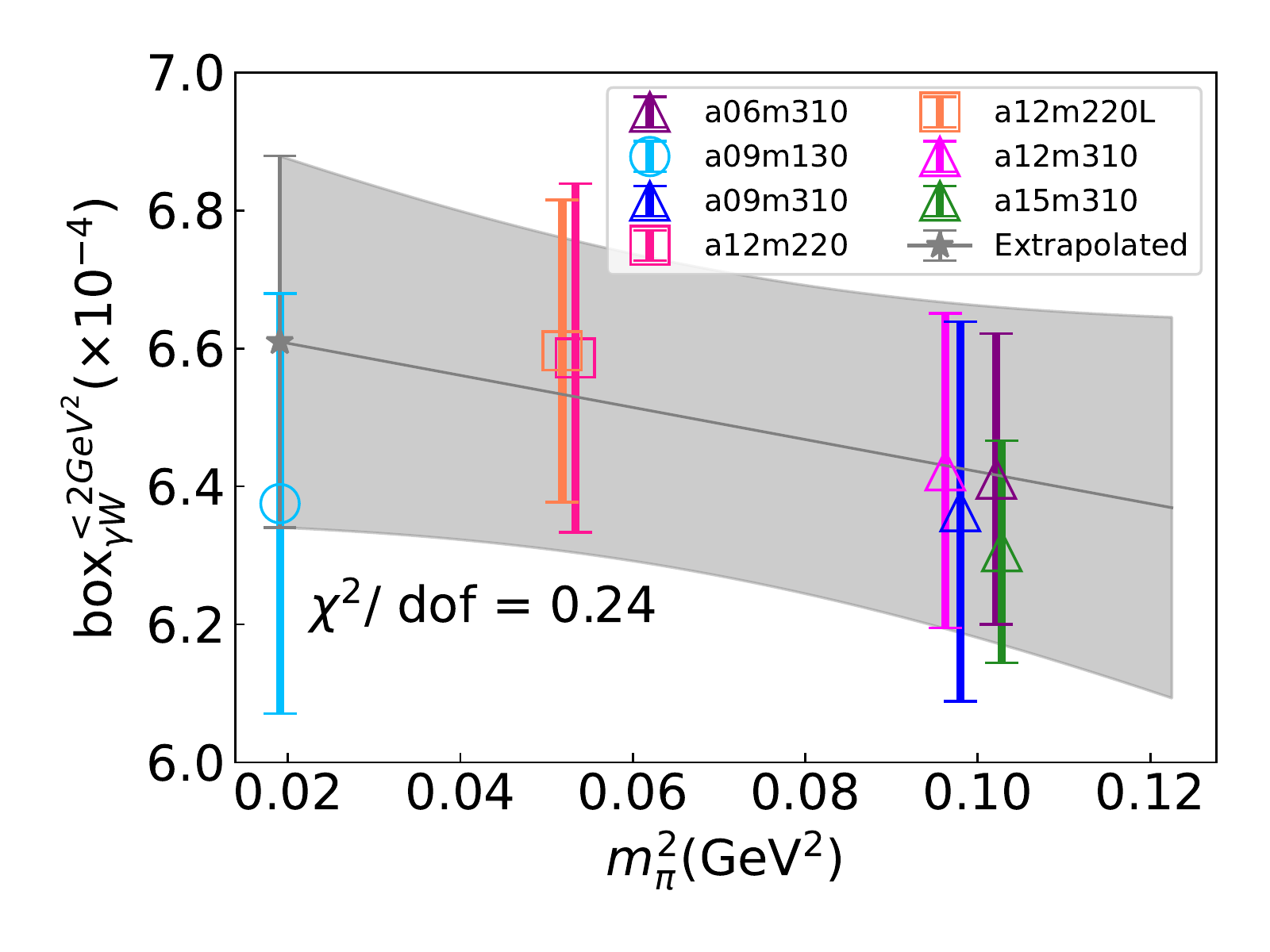}
    \label{fig:cont_chiral_extrapol_pi}
    \vspace{-0.7cm}
    \includegraphics[width=.5\textwidth]{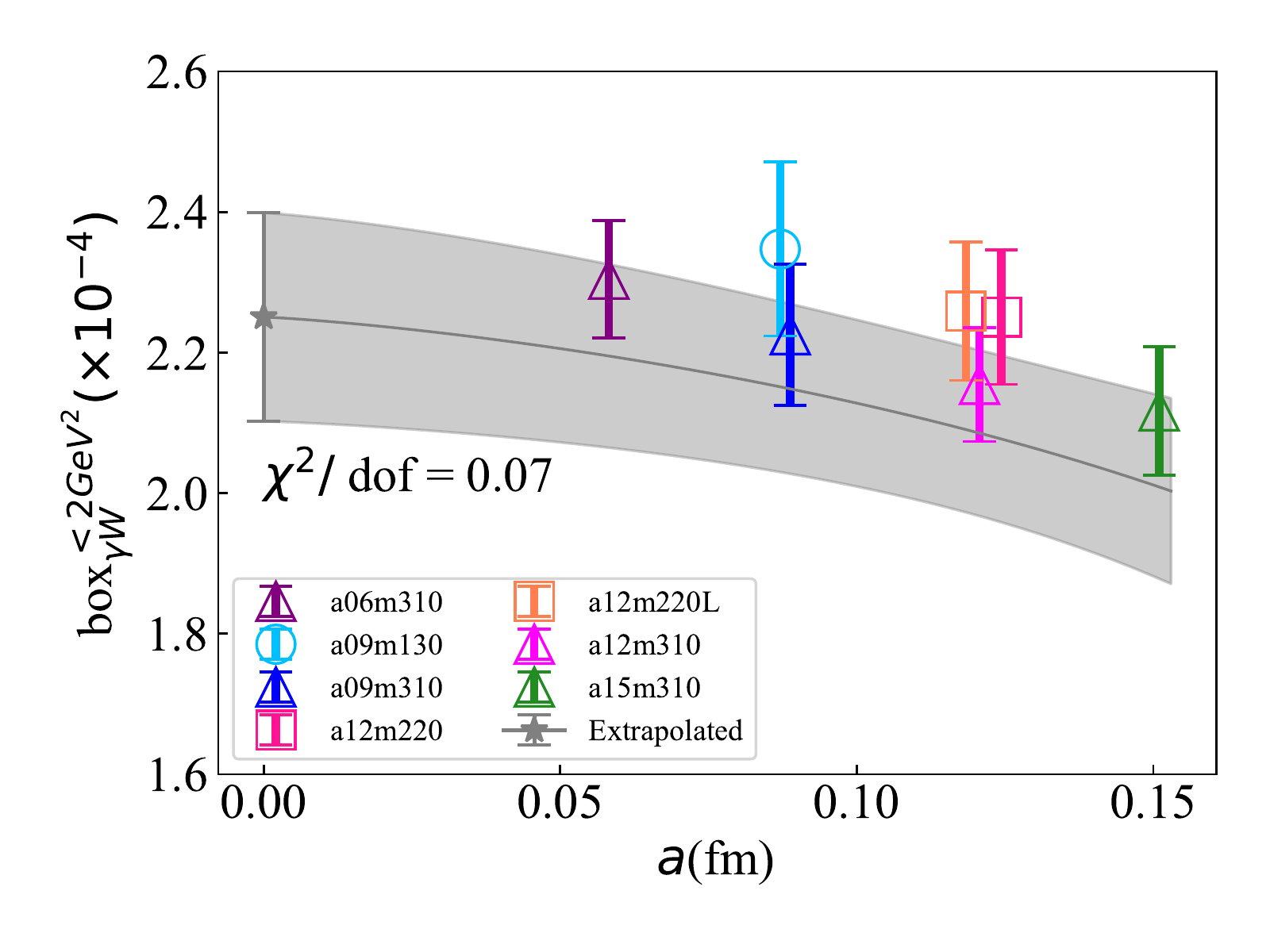}%
    \includegraphics[width=.5\textwidth]{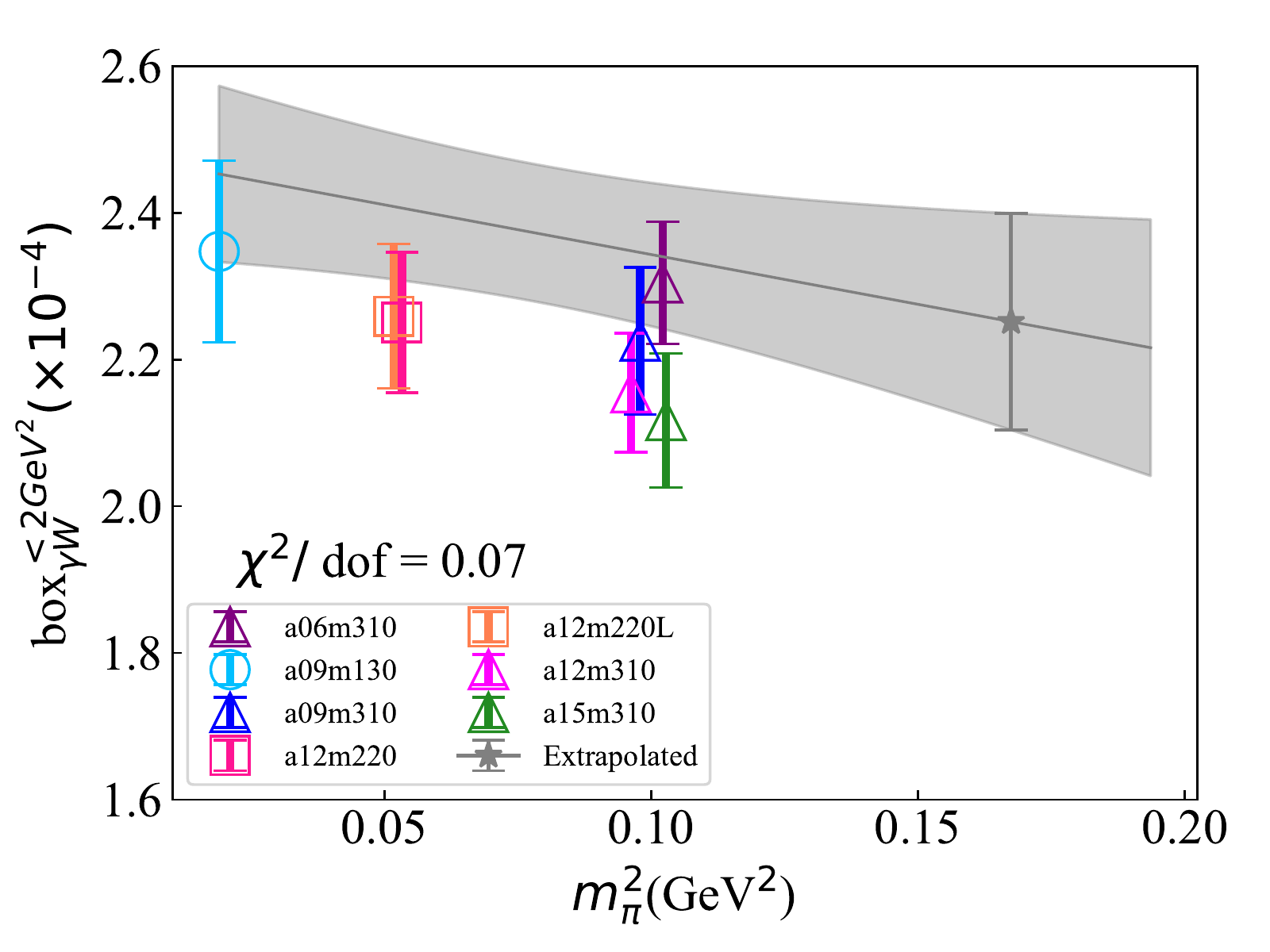}
    \vspace{-0.7cm}
    \caption{The dependence of the $\gamma W$-box contribution for $Q^2 \le 2 \textrm{GeV}^2$ for the pion (top) and kaon (bottom)  on  the lattice spacing $a$ (left), and  the pion mass ($M_\pi^2$) (right). The symbols for the various ensembles are defined in the inset and in Table~\ref{tab:my_label}. The results at the physical point are 
    shown by the grey star symbol. The result for the kaon is evaluated at the SU(3) symmetric point.}
    \label{fig:cont_chiral_extrapol_pi_K}
\end{figure}

As mentioned in Sec.~\ref{sec:EWbox},  ${\cal M}_H$ can be extracted
at all values of $Q^2$. In practice, we choose 60 $Q^2$ values that are the same on all 7 ensembles with a higher density below $Q^2 < 1$~GeV${}^2$.  These 60 points are converted into the smooth curves shown in Fig.~\ref{fig:M_H} (top) using a 
second-order interpolation. Data show that as $Q^2$ 
increases above $1$~GeV${}^2$, the value of ${\cal M}_H $ on coarser lattices 
decreases, indicating a dependence on the lattice spacing. Below $Q^2 < 1$~GeV${}^2$, 
the trend reverses. The integrated box contributions for $Q^2 < 2$~GeV${}^2$ and their dependence on $a$ and $M_\pi^2$ is shown in Fig.~\ref{fig:cont_chiral_extrapol_pi_K}. \looseness-1

\begin{figure}
    \centering
    \includegraphics[width=.5\textwidth]{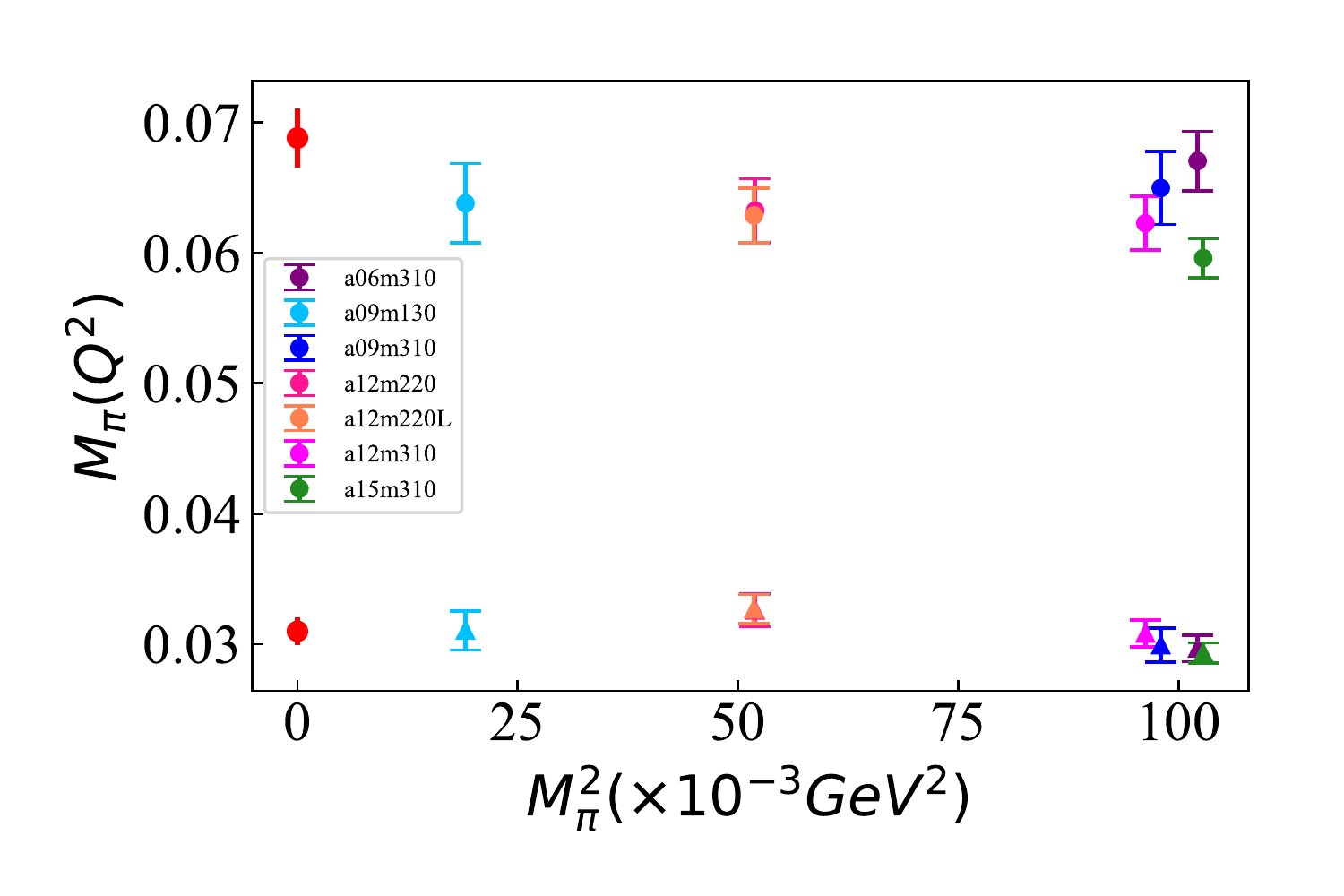}%
    \includegraphics[width=.5\textwidth]{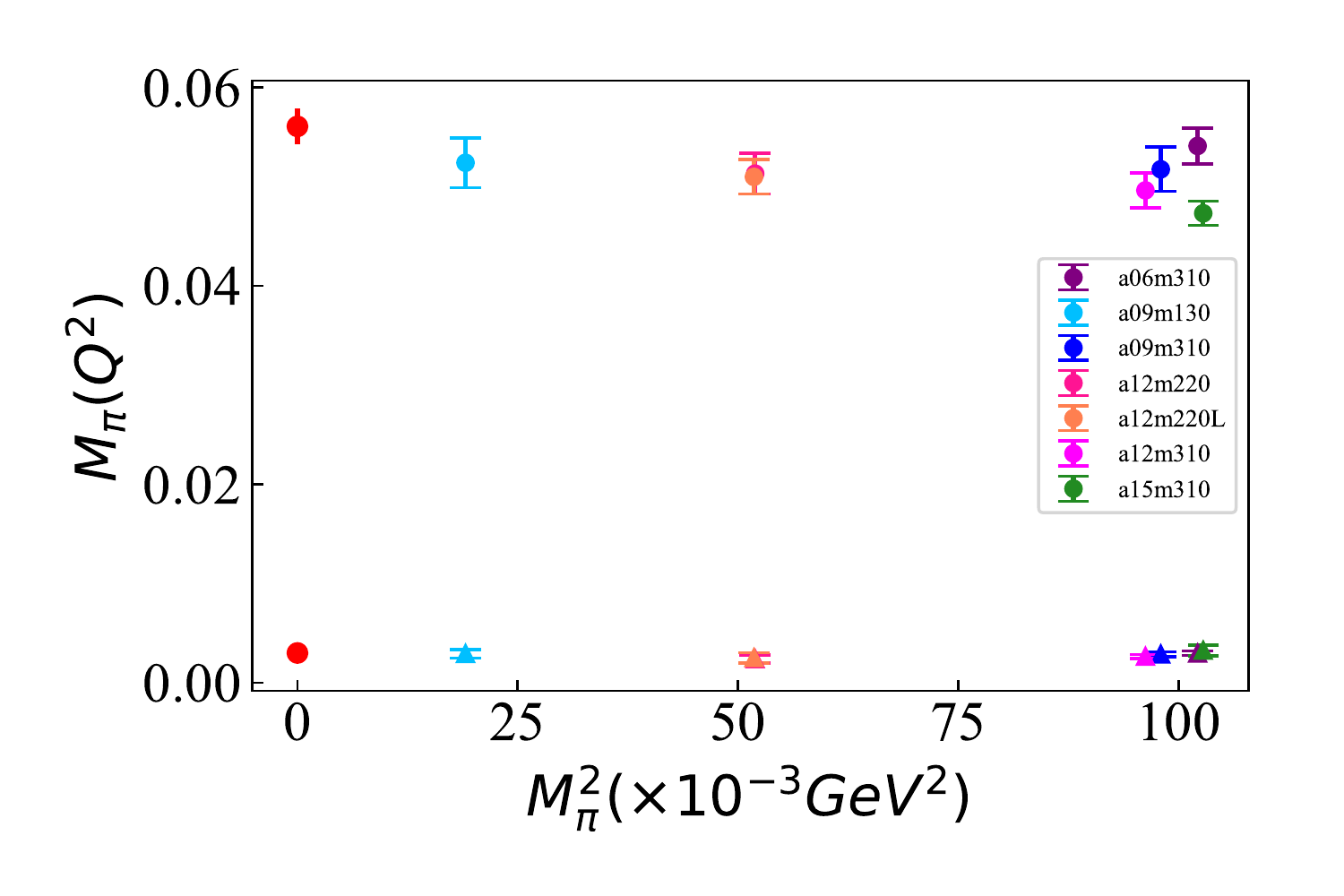}
    \caption{${\cal M}_H(Q^2)$ of (a) pion and (b) Kaon at $Q^2 = 0.133\textrm{ GeV}^2$ (triangle), $2.00\ \textrm{GeV}^2$ (circle). Ensembles are labeled by color. Data do not show a significant dependence of ${\cal M}_H (Q^2) $ on $M_\pi^2$. Red point is the continuum extrapolated value using a fit linear in $a\alpha_S$. }
    \label{fig:mass_dep}
\end{figure}

To compare the lattice ${\cal M}_H(Q^2)$ to perturbation theory, we extrapolate the data to the continuum limit at $M_\pi = 135$~MeV using a fit linear in $a \alpha_S$ since the dependence on $M_\pi$ is observed to be small (See Fig.~\ref{fig:mass_dep}). These fits, for all the ensembles and all $Q^2$ values, have a $p$-value above 0.2. 
As shown in Figure~\ref{fig:M_H}, this continuum limit data,   represented by the grey solid line, roughly agrees with perturbative result (gold line) for $Q^2 >  2 \textrm{GeV}^2$. Uncertainty in the perturbative result arises from the truncation 
(current result is $4^{\rm th}$ order) and the neglected higher-twist (HT) contributions~\cite{Feng:2020zdc}. Since diagram (A) only has HT contributions, we use its lattice value as an estimate of the HT uncertainty and show it by the dotted lines about the perturbative result.


\section{Continuum extrapolation of lattice data}
The extrapolation of the $\gamma W$-box for $Q^2 < Q_{\text{cut}}^2$ to the continuum limit $a=0$ and mass $M_\pi = M_\pi^{\text{phys}}$ for the pion, and $M_\pi = M_K^{\text{SU(3)}}$ for the kaon is carried out keeping the lowest order dependence on the pion mass ($M_\pi^2$) and on the lattice spacing ($\alpha_S a$):
\begin{equation}
    \Box |_{VA}^{Q^2 < Q_{\text{cut}}^2} ( M_\pi, a) = c_0 + c_1 a \alpha_S + c_2 M_\pi^2 \,.
\end{equation}
This extrapolation is shown in (Fig.~\ref{fig:cont_chiral_extrapol_pi_K}) and gives
    \begin{equation}
    \label{eq:box_pi_K}
        \square_{\gamma W}^{VA} |^{Q^2 \le 2 \textrm{GeV}^2}_{\pi}
        = 0.661 (27) \times 10^{-3} \,, \qquad
        \square_{\gamma W}^{VA} |^{Q^2 \le 2 \textrm{GeV}^2}_{K}
        = 0.225 (15) \times 10^{-3}
    \end{equation}
Systematic uncertainties due to the chiral-continuum extrapolation are included in these estimates. We also estimated possible uncertainty in ${\cal M}_H$ due to integration using 52 discrete points in $Q^2$ 
 as the difference between using the trapezoid and Simpson methods and found it to be negligible. We assume that finite volume effects are negligible  since all ensembles have $M_\pi L \ge 3.9$.

\section{Electroweak $\gamma W$-box correction and comparison to earlier work}

The contribution above the energy cut at $Q^2 = 2 \textrm{GeV}^2$ is computed using the operator product expansion~\cite{Feng:2020zdc} with the higher-twist uncertainty estimated using diagram A (See Fig.~\ref{subfig:diagrams}).
    \begin{equation}
    \label{eq:PT}
        \square_{\gamma W}^{VA}|_{\pi,K}^{Q^2 > 2 \textrm{GeV}^2} = 2.159(6)_{HO}(7)_{HT}\times 10^{-3}.
    \end{equation}
Combining Eq.~\eqref{eq:PT} with  Eq.~\eqref{eq:box_pi_K} gives our results for the full box contribution:
    \begin{align}
        \square_{\gamma W}^{VA} |_{\pi}
        = 2.820 (28) \times 10^{-3} \,, \qquad
        \square_{\gamma W}^{VA} \Big|_{K^{0, S U(3)}}
        = 2.384 (17) \times 10^{-3} \,,
    \end{align}
which are in good agreement with those from Feng et al.~\cite{Feng:2020zdc,Ma:2021azh}
    \begin{align}
         \square_{\gamma W}^{VA}|_{\pi} = 2.830(11)(26) \times 10^{-3} \,, \qquad
        \left.\square_{\gamma W}^{V A}\right|_{K^{0, S U(3)}} = 2.437(44) \times 10^{-3} \,.
    \end{align}
    The difference in $\square_{\gamma W}^{V A}\Big|_{K^{0, S U(3)}}$ is $1.2 \sigma$, but note that our value is determined with extrapolation in $M_\pi^2$ to $SU(3)-$symmetric point, while the Feng et al. value was computed at the physical pion mass, i.e., without extrapolation to $M_K|_{\rm SU(3)}$.
    Calculations for the nucleon are in progress.  
    
\vspace{0.2cm}
\noindent {\textbf {Acknowledgements}}: We thank the MILC collaboration for providing the HISQ lattices, and Vincenzo Cirigliano and Emanuele Mereghetti for discussions. The calculations used the CHROMA software suite~\cite{Edwards:2004sx}. Simulations were carried out at (i) the NERSC supported by DOE under Contract No. DE-AC02-05CH11231;  (ii) the USQCD collaboration resources funded by DOE HEP, and (iii) Institutional Computing at Los Alamos National Laboratory. This work was supported by LANL LDRD program and TB and RG were also supported by the DOE HEP  under Contract No. DE-AC52-06NA25396.
\bibliographystyle{JHEP}
\let\oldbibitem\bibitem
\def\bibitem#1\emph#2,{\oldbibitem#1}
\let\oldthebibliography\thebibliography
\renewcommand\thebibliography[1]{\oldthebibliography{#1}%
                                 \itemsep0pt\parskip0pt\relax}
\bibliography{main}

\end{document}